\documentclass[conference]{IEEEtran}

\usepackage{hyperref}
\usepackage{cite}

\usepackage{tikz}
\usepackage{pgfplots}

\usepackage{algorithm}
\usepackage{algpseudocode,balance}

\usepackage{amssymb}
\usepackage{amsmath}

\graphicspath{{./fig/}} 

\newcommand{\logtwo}{\mathop{\log_{2}}}
\newcommand{\logten}{\mathop{\log_{10}}}
\newcommand{\norm}[1]{\lVert#1\rVert}
\newcommand{\normzero}[1]{\norm{#1}_{0}}

\newcommand{\normtwosquare}[1]{\norm{#1}_2^2}
\newcommand{\normM}[1]{\norm{#1}_{2,0}}
\newcommand{\tr}[1]{\text{tr}(#1)}

\newcommand{\diag}[1]{\text{diag}\left(#1\right)}
\newcommand{\expect}[1]{\mathrm{E}[#1]}

\newcommand{\absqr}[1]{\lvert#1\rvert^2}

\IEEEoverridecommandlockouts
\title{Cloud Radio Access Networks With Coded Caching}
\author{ {Yi{\u{g}}it U{\u{g}}ur, Zohaib Hassan Awan and Aydin Sezgin}\\ \normalsize{Institute of Digital Communication Systems}\\ 	\normalsize{Ruhr-Universit\"{a}t Bochum, 44780, Germany}\\ \small{Email: \{yigit.ugur, zohaib.awan, aydin.sezgin\}@rub.de} }

\begin{document}
\maketitle

\begin{abstract}
A cloud radio access network (\mbox{C-RAN}) is considered as a candidate to meet the expectations of higher data rate demands in wireless networks. In \mbox{C-RAN}, low energy base stations (BSs) are deployed over a small geography and are allowed to connect to the cloud via finite capacity backhaul links where the information is processed. A conventional \mbox{C-RAN}, however, requires high capacity backhaul links, since the requested files need to be transferred first from the cloud to the BS before conveying them to the users. One approach to overcome the limitations of the backhaul links is to introduce local storage caches at the BSs, in which the popular files are stored locally in order to reduce the load of the backhaul links. Furthermore, we utilize coded caching with the goal to minimize the total network cost, i.e., the transmit power and the cost associated with the backhaul links. The initial formulation of the optimization problem for this model is non-convex. We first reformulate and then convexify the problem through some relaxation techniques. In comparison to the uncoded caching at the BSs, our results highlight the benefits associated with coded caching and show that it decreases the backhaul cost. 
\end{abstract}

\section{Introduction}
\label{sec:Introduction}

Mobile data traffic has experienced explosive growth in the last decade and is expected to increase further. Capabilities for supporting high data traffic and extremely low latency are some fundamental requirements identified for the future 5G mobile networks. A \mbox{C-RAN} model can be considered as a potential candidate to meet these requirements \cite{CRAN}. In \mbox{C-RAN}, due to economical reasons instead of a high power BS, several low power BSs are spread over a small area, which are allowed to connect to the  cloud via backhaul links. High speed fiber cables are generally used to establish backhaul links for low latency. One drawback is that establishing all the links via fiber optical cables from the cloud to the BSs may lead to high infrastructure cost. A possible alternative to save on cost is by utilizing finite capacity wireless backhaul links. However, this may lead to higher latency. A potential solution to solve this issue is by introducing local storage or caches at the BSs~\cite{golrezaei}, where the most popular contents, for example, files, movies, or multimedia, are stored, locally. When a user requests a file which is available at the cache of the BS, it can be directly downloaded from the BS.  In other words, this approach reduces the load of backhaul links with relatively inexpensive local storage.

In general, the cache placement schemes can be roughly partitioned into two types, namely, uncoded and coded caching~\cite{shanmugam,caching}. In uncoded caching, BSs store the complete file; while, in coded caching,  instead of complete file, small fractions of the files are stored at the caches using fountain or maximum distance separable (MDS) codes~\cite{shanmugam},~\cite{liu},~\cite{liu2}. The main advantage of the coded caching compared to the uncoded caching is that the probability of re-constructing the file at the desired receiver without using the backhaul links is higher. Subsequently, it reduces the backhaul cost. In addition to this, if the users still need to use backhaul links to recover the files --- for example, when the portion of the file required by the users are not available in the cache --- only remaining portion of the files will be transferred from the cloud~\cite{caching}; and, under secrecy constraints in~\cite{zohaib}. In~\cite{deneme}, minimizing the backhaul and transmission power cost is considered where uncoded caching is used at the BSs. This optimization problem is formalized with the help of semidefinite programming (SDP) relaxation and $l_{0}$-norm approximation. In~\cite{yuanming,shixin}, the authors studied a C-RAN in the absence of caching and study the problem of resource allocation.  

In this work, we consider a C-RAN model as shown in Fig.~\ref{fig:architecture}, where the BSs are equipped with multiple antennas. The users can download the files either locally from the BSs or from the central storage at the cloud via backhaul links. The choice of beamforming vectors and how to  appropriately fill the caches are decided in the central processor at the cloud. The main goal of this work is to minimize the downlink network cost of the aforementioned C-RAN system. We refer to the downlink network cost as the sum of the total transmission power and the backhaul cost. We optimize the downlink cost over beamforming vectors such that each user has a minimum quality of service (QoS). In this work, we consider the case where coded caching at the BSs is used. Users that request the file, receive multiple fractions of the file from different caches of the BSs and try to reconstruct the desired file. The initial optimization problem consists of a non-convex indicator function and a parameter that models the missing fractions of the files at the serving BSs. We first reformulate the original problem by using a maximum function, then solve it with the help of SDP relaxation. The simulation results show that the coded caching  outperforms the uncoded caching strategies in a variety of situations.

\IEEEpubidadjcol 

We structure this paper as follows. The system model is described in Section \ref{sec:SystemModel}. The optimization problem is formally presented in Section \ref{sec:Optimization}. Simulation results are presented in Section \ref{sec:Result}. Finally, in Section \ref{sec:Conclusion}, we conclude the paper by summarizing its contributions.

\vspace{0.5em}
\emph{Notation:} Bold small and capital letters denote vector and matrices, respectively. The superscripts $(\cdot)^{\text{H}}$ and $(\cdot)^{-1}$ represent the Hermitian transpose and the matrix inverse, respectively. The function $\{x\}^+$ denotes the maximum value between 0 and $x$. The $l_{p}$-norm is denoted by $\norm{\cdot}_{p}$. The expectation is denoted by $\expect{\cdot}$, the trace operation is represented by $\mathrm{tr}(\cdot)$ and  $\mathbb{C}^{x\times y}$ denotes the space of $x\times y$ complex matrices. For a square matrix $\mathbf{A}$, we use $\mathbf{A} \succeq \mathbf{0}$ to indicate that $\mathbf{A}$ is positive semidefinite. The zero vector of size $N \times 1$ is represented by $\mathbf{0}_{N}$.

\section{System Model and Problem Formulation}
\label{sec:SystemModel}
\begin{figure}
\begin{tikzpicture}[scale=0.5335]
	
\node (cloud) at
(0,9){\includegraphics[width=3.8cm,height=1.6cm]{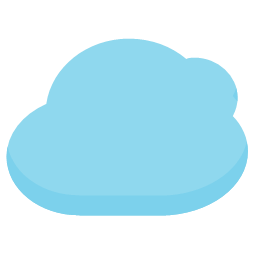}};
	
\node (bs1) at (-3,2){\includegraphics[scale=.07]{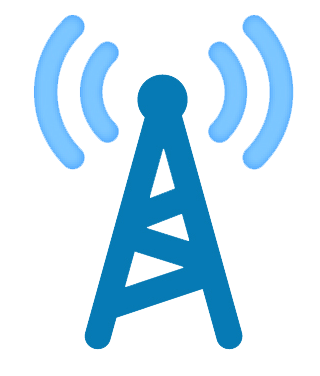}};
\node (bs2) at (0,-2.2){\includegraphics[scale=.07]{BS.png}};
\node (bs3) at (3,2){\includegraphics[scale=.07]{BS.png}};
	
\node (storage1) at (-3,0.9){\includegraphics[scale=.08]{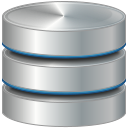}};
\node (storage2) at (0,-3.3){\includegraphics[scale=.08]{STORAGE.png}};
\node (storage3) at (3,0.9){\includegraphics[scale=.08]{STORAGE.png}};
		
\node (mu1) at (-6,1){\includegraphics[scale=.002]{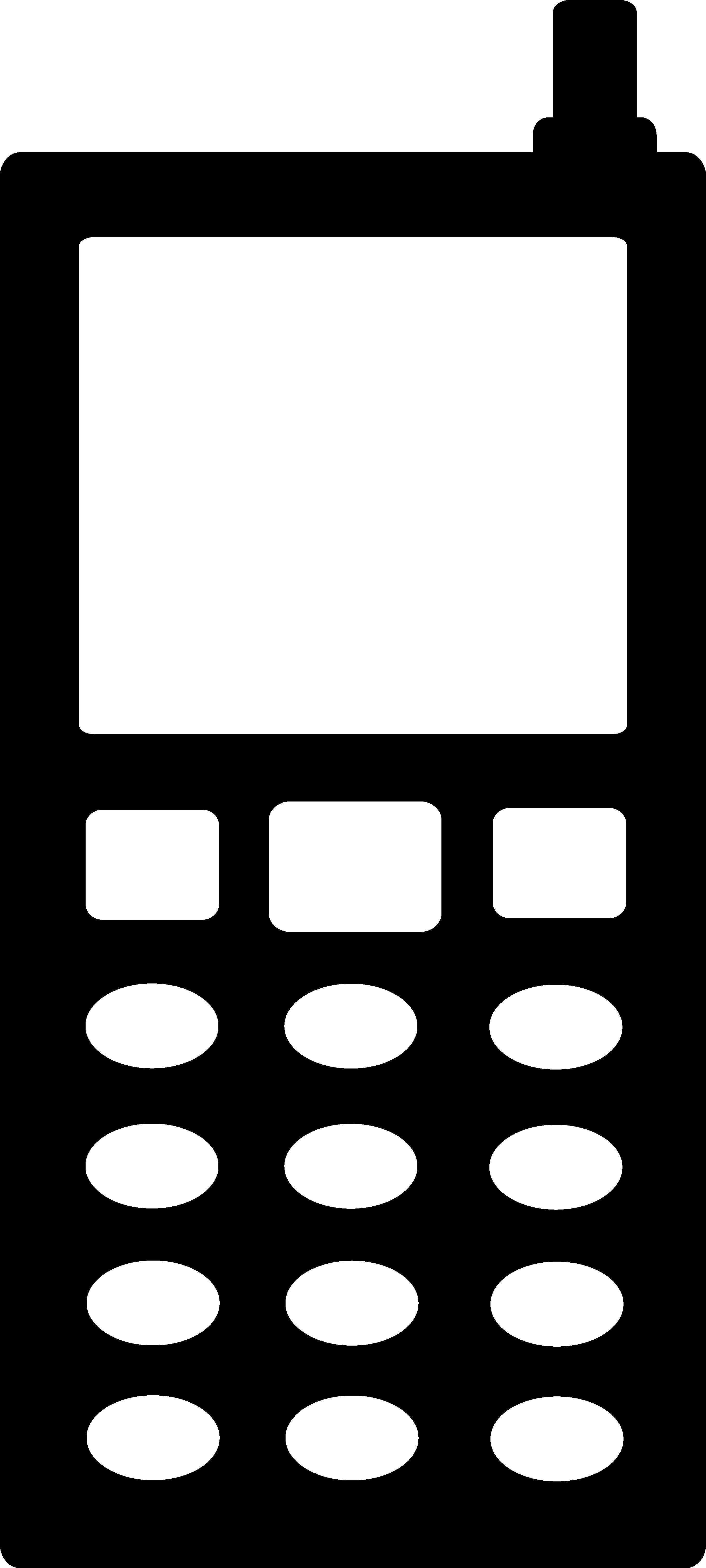}};
\node (mu2) at (-1.5,0.5){\includegraphics[scale=.002]{MU.png}};
\node (mu3) at (2,-3){\includegraphics[scale=.002]{MU.png}};
\node (mu4) at (-5,-0.5){\includegraphics[scale=.002]{MU.png}};
\node (mu5) at (-3,-2){\includegraphics[scale=.002]{MU.png}};
\node (mu6) at (-1.5,0.5){\includegraphics[scale=.002]{MU.png}};
\node (mu7) at (1,3){\includegraphics[scale=.002]{MU.png}};
\node (mu8) at (1.5,-0.5){\includegraphics[scale=.002]{MU.png}};
\node (mu9) at (3.5,-2){\includegraphics[scale=.002]{MU.png}};
\node (mu10) at (5.5,-0.5){\includegraphics[scale=.002]{MU.png}};
\node (mu11) at (5,1.5){\includegraphics[scale=.002]{MU.png}};
	
\draw [green,thick,dashed] (0,0) ellipse (7cm and 4cm);
\draw [red,thick] (0,6) ellipse (1.8cm and 0.5cm);
	
\draw [red, thick, double, line cap=round] (-0.5,7.5) -- (bs1);
\draw [red, thick, double, line cap=round] (0,7.5) -- (bs2);
\draw [red, thick, double, line cap=round] (0.5,7.5) -- (bs3);
	
\node at (-0.1,9) {Cloud};
\node[red] at (4,6) {Backhaul Links};
	
\draw[thick] (-7,7.7) rectangle (-2.5,4.5);  	
	
\node at (-4.2,7) {Base Station};
\node (bs) at (-6.5,7){\includegraphics[scale=.05]{BS.png}};
	
\node at (-5.2,6) {User};
\node (bs) at (-6.5,6){\includegraphics[scale=.002]{MU.png}};
	
\node at (-5,5) {Cache};
\node (bs) at (-6.5,5){\includegraphics[scale=.08]{STORAGE.png}};
\end{tikzpicture}
\centering
\caption{System architecture of a C-RAN with local cache storage.}
\label{fig:architecture}
\end{figure}

We consider a C-RAN composed of $L$ BSs and $M$ users. Each BS is equipped with $N_{t}$ transmit antennas and each user has a single antenna as shown in Fig.~\ref{fig:architecture}. All the BSs are connected to the cloud via finite capacity backhaul links; and, there are no direct links between the users and the central cloud storage. In addition to the central file storage, local storage called cache are available at each BS. The caches have a smaller storage capacity compared to the cloud and are thus able to store only a limited portion of files. We focus on the downlink phase of C-RAN in which the users request files. The files can be delivered in two ways, 1) direct transmission from the BSs --- if they are available locally in their cache,  2) by prefetching the files from the cloud to the BSs and then transmitting it to the users.

The channel vector from the $l$-th BS to the $m$-th user is denoted by ${\mathbf{h}_{l,m}\in\mathbb{C}^{N_{t}\times1}}$. The overall channel vector from all the BSs to the $m$-th user can be concisely given by ${\mathbf{h}_{m}=[\mathbf{h}_{1,m}^{\text{H}}, \mathbf{h}_{2,m}^{\text{H}}, \hdots, \mathbf{h}_{L,m}^{\text{H}}]^{\text{H}} \in \mathbb{C}^{L N_{t}\times 1}}$. We assume that the channel is constant and do not change in a time slot. The channel state information is known to all nodes, globally. The received signal at the $m$-th user can be expressed as
\begin{multline}
 y_{m}(t)  = \mathbf{h}_{m}^{\text{H}} \mathbf{w}_{m}(t) s_{m}(t) + \sum\limits_{j \neq m}^M \mathbf{h}_{m}^{\text{H}}\mathbf{w}_{j}(t) s_{j}(t) + z_{m}(t),\\\quad m=1,\hdots, M, \quad 1\leq t \leq T  
\label{eq:SigRx}
\end{multline}
where ${\mathbf{w}_{m}(t) =[\mathbf{w}_{1,m}^{\text{H}}(t), \mathbf{w}_{2,m}^{\text{H}}(t), \hdots, \mathbf{w}_{L,m}^{\text{H}}(t)]^{\text{H}} \in \mathbb{C}^{L N_{t}\times 1}}$ is the beamforming vector from all the BSs to the $m$-th user, $\mathbf{w}_{l,m}(t)$ is the beamforming vector from the $l$-th BS to the $m$-th user, $s_{m}(t)$ is the desired data symbol for the $m$-th user and $z_{m}(t)$ denotes the independent identically distributed additive complex Gaussian noise, with zero mean and variance $\sigma^2$ at users, $\forall$ $m$. Note that $T$ is the number of channel uses needed to fulfill the demands of the users. For convenience, we skip the time index $t$ in the rest of the paper. Without loss of generality, we assume that $\expect{\absqr{s_{m}}}=1$ and $s_{m}$'s are chosen independently from each other. By treating the contribution of undesired signals at the $m$-th user as noise, the corresponding signal-to-interference-plus-noise ratio (SINR) can be written as
\begin{align}
\text{SINR}_{m}=\frac{\absqr{\mathbf{h}_{m}^{\text{H}} \mathbf{w}_{m}}}{
\sum_{j \neq m}^M \absqr{\mathbf{h}_{m}^{\text{H}}
\mathbf{w}_{j}} + \sigma^2}, \quad m=1,\hdots, M.
\label{eq:SINR}
\end{align}
We assume that each user has a minimum SINR requirement that needs to be fulfilled to satisfy a particular quality of service constraint. We refer to the target SINR for the $m$-th user as $\gamma_{m}$.

\subsection{Files Placement in the Caches}
\label{sec:Cache}
Since the storage capacity of caches is lower than that of the central cloud, how to fill them is of paramount importance. The main idea is to only store the popular contents in the caches, so as to avoid the data transfer cost that arises from the backhaul link. The popularity of the files can be measured based on the number and behavior of the requests; and, is modeled here as Zipf distribution~\cite{golrezaei}. We assume that the files are sorted from the most to the least popular ones, such that the most popular file has index 1 ($f=1$) and the least popular file has index $F$ ($f=F$). According to Zipf distribution, the probability of $f$-th file requested by the user is given as
\begin{align}
Z_{f}=\frac{f^{-\alpha}}{\sum_{k=1}^{F}k^{-\alpha}}, \quad
f=1,\hdots, F
\end{align}
where $\alpha$ is the shaping parameter and is related to the skewness of the distribution \cite{zipf}. A large value  of $\alpha$ means that the distribution is more skewed and therefore the probability of requesting a small group of files is large. We assume that the popularity distribution is known at the cloud.

We define the cache placement matrix  $\mathbf{R} \in [0,1]^{F \times L}$, where the $f$-th row and the $l$-th column of $\mathbf{R}$ is $\delta_{f,l}$, which corresponds to the fraction of the parity bits of the $f$-th file that is stored in the cache of the $l$-th BS. We denote the storage capacity of each cache as $BS$ bits, where $B$ is the size of each file. Due to the cache size constraint,
\begin{align}
\sum\nolimits_{f=1}^{F}\delta_{f,l}=BS
\end{align}
needs to be satisfied at the $l$-th BS, $\forall$\: $l$ \footnote{For convenience, in the rest of the paper, we assume that all files have normalized size ($B=1$).}. The cache placement matrix $\mathbf{R}$ is fixed. This is due to the fact that, the popularity of files changes slowly in comparison to the channel variations. As mentioned before, MDS or fountain codes can be used to encode files which are then stored in the caches of the BSs. By using an ideal MDS code, a file is encoded into an arbitrary long number of parity bits, such that after recovering any of $B$ parity bits, the receiver is able to recover the desired message. We assume that the cache at each BS stores a unique subset of parity bits. Then, the $m$-th user can download the $f$-th file if the following condition is fulfilled
\begin{eqnarray}
\label{eq:cond1}
\sum\nolimits_{l \in \mathcal{H}(m)} \delta_{f,l} \geq 1
\end{eqnarray}
where $\mathcal{H}(m)$ is the set of the BSs that serve the $m$-th user. If  all elements of the beamforming vector from the $l$-th BS to the $m$-th user equals zero ($\mathbf{w}_{l,m}=\mathbf{0}_{N_{t}}$), i.e., the input power of beamforming vector $\mathbf{w}_{l,m}$ is zero, the $l$-th BS does not serve the $m$-th user. Subsequently,~\eqref{eq:cond1} can be rewritten  as
\begin{eqnarray}
\label{eq:cond2}
\sum\nolimits_{l=1}^L \delta_{f,l} \normzero{
\normtwosquare{\mathbf{w}_{l,m}}} \geq 1
\end{eqnarray}
where $l_{0}$-norm gives the number of nonzero entries in a vector. For notational simplicity, in the rest of the paper, we will use $\normM{\cdot}$ to replace  $\normzero{\normtwosquare{\cdot}}$.

\subsection{Network Cost Model}
We define the total network cost of the C-RAN as the sum of backhaul and power consumption cost. The overall network cost $C_{\text{N}}$ can be written as
\begin{eqnarray}
\label{network-cost}
C_{\text{N}}=\lambda C_{\text{BH}} + (1-\lambda) C_{\text{P}}
\end{eqnarray}
where $\text{C}_{\text{BH}}$ and $\text{C}_{\text{P}}$ denote the backhaul cost and the power consumption cost, respectively. The parameter $\lambda$ indicates the relative emphasis on the backhaul cost and the power consumption cost, where $\lambda \in [0,1]$. Starting from~\eqref{network-cost}, the overall network cost can be written as
\begin{eqnarray}
\label{main-problem}
C_{\text{N}} =\lambda  \underbrace{\sum\limits_{m=1}^M \sum\limits_{f=1}^F Z_{f}\mathbf{1}_{m,f}C_{\text{BH}}^{(m,f)}}_{:=C_{\text{BH}}} + (1-\lambda)  \underbrace{\sum\limits_{m=1}^M \normtwosquare{\mathbf{w}_{m}}}_{:=C_{\text{P}}}
\end{eqnarray}
where $\mathbf{1}_{m,f}$ is the indicator function and is given by
\begin{align}
\mathbf{1}_{m,f}= \left\{
\begin{array}{ll}
0, \quad & \quad \sum\limits_{l=1}^L \delta_{f,l} \normM{\mathbf{w}_{l,m}}  \geq1\\
1, \quad & \quad \text{otherwise}
\end{array} \right.
\end{align}
and $C_{\text{BH}}^{(m,f)}$ is the associated backhaul cost for the case in which the $f$-th file required by the $m$-th user can not be provided by the BSs. We define $C_{\text{BH}}^{(m,f)}$ as
\begin{eqnarray}
C_{\text{BH}}^{(m,f)}=X_{m,f} R_{m}
\end{eqnarray}
where $X_{m,f}$ denotes the missing portion of the $f$-th file not present in the serving caches of the $m$-th user. This fraction of the $f$-th file needs to be transfered via backhaul. The transmission rate $R_{m}$ for the $m$-th user is given by  $R_{m}=\logtwo(1+\gamma_{m})$.

\section{Optimization Problem}
\label{sec:Optimization}
In this section, we study the network cost problem stated in~\eqref{main-problem}. The goal is to minimize the total network cost and in doing so,  we need to optimize  the beamforming vectors such that 1) the quality of service constraints (in this case  minimum SINR targets) of each user is fulfilled, and 2) the maximum transmit power constraints at each BS is met. Thus, we can formulate our optimization problem as follows
\begin{subequations}
\label{problem-1}
\begin{align}
\hspace{-0.5em}\min_{\mathbf{w}_{m}} \hspace{1em} &\lambda
\sum\limits_{m=1}^M \sum\limits_{f=1}^F Z_{f}\mathbf{1}_{m,f}X_{m,f}R_{m} + (1-\lambda) \sum\limits_{m=1}^M \normtwosquare{\mathbf{w}_{m}} \label{obj}\\  \vspace{0.7em}\hspace{-5em}\text{s. t.}\:\:\: & \text{SINR}_{m} \geq \gamma_{m},\quad \forall m \label{constraint1} \\ \vspace{0.15em}
&\sum\limits_{m=1}^M \normtwosquare{\mathbf{w}_{l,m}} \leq P_{\text{max}},  \quad \forall l \label{constraint2}
\end{align}
\end{subequations}
where $P_{\text{max}}$ is the maximum transmission power of each BS. Note that the indicator function makes the objective non-convex. Starting from~\eqref{problem-1}, the optimization problem can be identically expressed as
\begin{subequations}
\label{problem-2}
\begin{align}
\hspace{-5em}\min_{\mathbf{w}_{m} }  \hspace{1.5em}&\lambda
\sum\limits_{m=1}^M \sum\limits_{f=1}^F Z_{f}\bigg
\{1-\sum\limits_{l=1}^L \delta_{f,l} \normM{\mathbf{w}_{l,m}}
\bigg\}^{+} R_{m} \notag\\\hspace{-2.25em}  &\:\:\:\:+ (1-\lambda)\sum\limits_{m=1}^M \normtwosquare{\mathbf{w}_{m}} \label{obj2} \\ \vspace{0.7em}
\text{s. t.}\:\:\: & \eqref{constraint1}, \eqref{constraint2}.
\end{align}
\end{subequations}
In~\eqref{problem-2}, the parameter $X_{m,f}$ and the indicator function $\mathbf{1}_{m,f}$ from~\eqref{obj} are condensed into the function $\{\cdot\}^+$. Next, we reformulate  the objective function~\eqref{obj2} by introducing an auxiliary parameter $\beta_{f,m}$, as
\begin{subequations}
\label{problem-3}
\begin{align}
\hspace{-0.5em}\min_{\mathbf{w}_{m},\beta_{f,m} } \hspace{1em} &\lambda
\sum\limits_{m=1}^M \sum\limits_{f=1}^F Z_{f} \beta_{f,m} R_{m} +
(1-\lambda) \sum\limits_{m=1}^M \normtwosquare{\mathbf{w}_{m}} \\
\vspace{0.7em}
\text{s. t.}\:\:\: & \beta_{f,m} \geq 0, \quad \forall f, \forall m \\
\vspace{0.15em}
& 1-\sum\limits_{l=1}^L \delta_{f,l} \normM{\mathbf{w}_{l,m}} \leq
\beta_{f,m}, \quad \forall f, \forall m\label{problem-3c} \\ \vspace{0.15em}
& \eqref{constraint1}, \eqref{constraint2}. 
\end{align}
\end{subequations}
Due to the presence of quadratic terms in~\eqref{eq:SINR}, the SINR constraints in~\eqref{constraint1} render the problem non-convex. We solve this issue by semidefinite programming relaxation to convexify the SINR constraints. Let $\mathbf{W}_{m}$ and $\mathbf{H}_{m}$ be given by
\begin{eqnarray}
\mathbf{W}_{m}= \mathbf{w}_{m} \mathbf{w}_{m}^{\text{H}} \\
\mathbf{H}_{m}= \mathbf{h}_{m} \mathbf{h}_{m}^{\text{H}}
\end{eqnarray}
where $\mathbf{W}_{m}$ and $\mathbf{H}_{m}$ are positive semidefinite matrices. Then, $\normtwosquare {\mathbf{w}_{l,m}}$ in~\eqref{eq:cond2} and~\eqref{problem-3c} can be concisely written as
\begin{eqnarray}
\normtwosquare {\mathbf{w}_{l,m}}=\tr{\mathbf{W}_{m} \mathbf{J}_{l}}
\end{eqnarray}
where $\mathbf{J}_{l}$ is selection matrix~\cite{deneme} such that
\begin{eqnarray}
\mathbf{J}_{l}=\diag{\left[ \mathbf{0}_{(l-1)N_{t}}^{\text{H}},
\mathbf{1}_{N_{t}}^{\text{H}}, \mathbf{0}_{(L-l)N_{t}}^{\text{H}} \right]}.
\end{eqnarray}
After the aforementioned SDP relaxation, the problem is still non-convex because of the presence of the $l_{0}$-norm in the constraints \eqref{problem-3c}. The $l_{0}$-norm can be approximated by the logarithmic function \cite{norm} as follows
\begin{eqnarray}
\normzero{\mathbf{X}}= \log(\tr{\mathbf{X}}+\theta)
\end{eqnarray}
where $\theta$ is a parameter to adjust the smoothness, \mbox{$0 < \theta \leq 1$}. Finally, after the SDP relaxation the optimization problem in~\eqref{problem-3} can be written as
\begin{subequations}
\label{problem-4}
\begin{align}
\hspace{-0.5em}\min_{\mathbf{W}_{m},\beta_{f,m} }\hspace{0.1em}&\lambda
\sum\limits_{m=1}^M \sum\limits_{f=1}^F Z_{f} \beta_{f,m} R_{m} +
(1-\lambda) \sum\limits_{m=1}^M \tr{\mathbf{W}_{m}}\\  \vspace{0.7em}\:\:\:\:\:\:\: \text{s. t.} & \nonumber\\
&\hspace{-2.5em}\beta_{f,m} \geq 0, \quad \forall f, \forall m \\ \vspace{0.15em}
&\hspace{-2.5em} 1-\sum\limits_{l=1}^L \delta_{f,l}
\log\left(\tr{\mathbf{W}_{m} \mathbf{J}_{l}}+\theta\right)  \leq
\beta_{f,m}, \quad \forall f, \forall m\\  \vspace{0.15em}
&\hspace{-2.5em}  \frac{\tr{\mathbf{W}_{m}
\mathbf{H}_{k}}}{\sum\limits_{n \neq m}^M \tr{\mathbf{W}_{n}
\mathbf{H}_{k}} + \sigma^2} \geq \gamma_{m} \quad \forall m, \forall k \\  \vspace{0.15em}
&\hspace{-2.5em} \sum\limits_{m=1}^M \tr{\mathbf{W}_{m} \mathbf{J}_{l}}
\leq P_{\text{max}}, \quad \forall l \\  \vspace{0.15em}
&\hspace{-2.5em}  \mathbf{W}_{m} \succeq \mathbf{0}, \quad \forall m.
\end{align}
\end{subequations} 
\label{constaint3}
\noindent Let $\mathbf{W}_{m}^{*}$ denote the optimum solution of~\eqref{problem-4}. If $\mathbf{W}_{m}^{*}$ is rank-one, we can readily apply eigenvalue decomposition to get the optimum beamforming vectors. Otherwise, randomization methods~\cite{randomization,randomization2} can be used to obtain the beamforming vectors from the optimum solution $\mathbf{W}_{m}^{*}$. Algorithm \ref{algorithmR} concisely captures the steps for the randomization methods.

\begin{algorithm} [t!]
\begin{algorithmic}[1]
\State Initialize the minimum objective value
\Repeat
\State Generate random vectors $\mathbf{w}_{m}$ according to complex
Gaussian distributions $\mathbf{w}_{m} \sim \mathcal{N}(\mathbf{0},
\mathbf{W}_{m}^{*})$
\State Check whether constraints (\ref{constraint1}) and
(\ref{constraint2}) satisfies
\If{all constraints satisfy}
\State Plug in the $\mathbf{w}_{m}$ into the objective function (\ref{obj})
\State Compare the current minimum objective value to the previous one and retain the smaller one
\EndIf
\Until{number of iterations are reached}
\end{algorithmic}
\caption{Randomization for~\eqref{problem-4} }
\label{algorithmR}	
\end{algorithm}

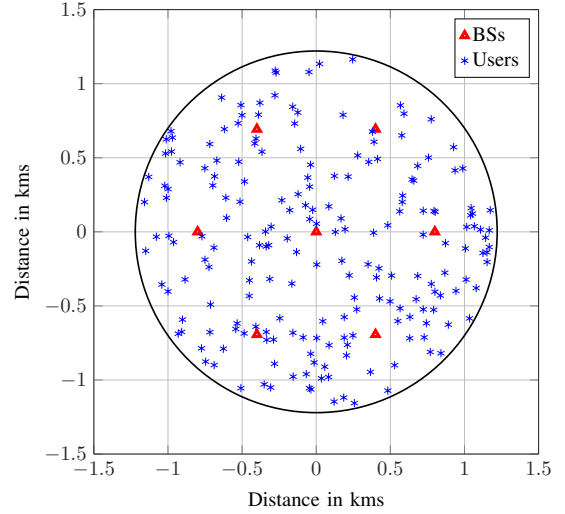
\begin{figure}
\centering
\definecolor{mycolor1}{rgb}{0.00000,0.44700,0.74100}
\definecolor{mycolor2}{rgb}{0.85000,0.32500,0.09800}
\begin{tikzpicture}[scale=0.776]
	
\begin{axis}[
width=3in,
height=3in,
scale only axis,
separate axis lines,
every outer x axis line/.append style={white!15!black},
every x tick label/.append style={font=\color{white!15!black}},
xmin=-1.5,
xmax=1.5,
xlabel={Distance in kms},
xmajorgrids,
every outer y axis line/.append style={white!15!black},
every y tick label/.append style={font=\color{white!15!black}},
ymin=-1.5,
ymax=1.5,
ylabel={Distance in kms},
ymajorgrids,
legend style={draw=black,fill=white,legend cell align=left} ]
	
\addplot[color=red,line width=1.5pt,only marks,mark=triangle,mark options={solid}]
table[row sep=crcr] {
0	0\\
0.4	0.692820323027551\\
-0.4	0.692820323027551\\
-0.4	-0.692820323027551\\
0.4	-0.692820323027551\\
-0.8	0\\
0.8	0\\ };
\addlegendentry{BSs};
	
\addplot[color=blue,only marks,mark=asterisk,mark options={solid}]
table[row sep=crcr]{
-0.164453674517865	-0.349693378268157\\
-0.885036096615534	-0.32148494986079\\
1.06633356342758	0.0372007890153988\\
-0.452012835554403	-0.433075376359689\\
0.942556085388892	-0.399189175794902\\
-0.697683222584015	0.312781226376392\\
0.426714687349015	-0.449970100484033\\
1.07878032730128	-0.37969199028369\\
-0.382685322943151	-0.0902141994364898\\
-0.351153734184184	-1.02964441911753\\
0.658153559784622	0.345880391976177\\
-0.969823109545487	0.635706846701565\\
-0.287969164119034	-0.727625691154887\\
0.0033725629671223	-0.716873637691513\\
0.0849276820644755	0.171544727234719\\
-0.0370977729252719	-0.823496317617647\\
1.15155662572512	-0.138306029637192\\
-0.133081021612126	-0.136772148677383\\
-0.774152467665706	-0.78637508327459\\
1.16690990719559	-0.102886171081882\\
0.0346918686240572	-0.989184405483214\\
0.800551769860469	-0.404474274462736\\
0.684438876412709	-0.297797074822454\\
-0.723234323494694	-0.237779930212744\\
0.167392686479503	0.0901848201841146\\
0.632006219383419	-0.574509662099935\\
0.926652440867984	0.571391490470302\\
-1.02162109974928	0.311117317110069\\
0.673863005971575	-0.45482316861221\\
0.583061636692488	0.244535556339586\\
-1.07843651472813	-0.0331458293103309\\
-0.320497419358638	-0.089989329802327\\
-0.918331221061105	0.469222389616648\\
0.353885921774746	0.471616297618778\\
0.721272322752324	-0.0189731804493842\\
-1.01158480577682	0.624703420895674\\
1.16700322936776	0.0100034498833094\\
0.93623385561332	0.414843139810741\\
0.988457600123428	0.427984344461193\\
-1.04142170631022	-0.355741107790226\\
-0.0442041745780144	0.304441972619014\\
0.591767656748367	0.797439500335592\\
0.742991099558358	-0.617106783119439\\
-0.304304476822997	0.034616227140413\\
-0.407200503898863	-0.639865848651036\\
0.794051799579132	0.14070379404899\\
-0.178802467655002	0.146245191837579\\
-0.962409491398053	-0.0697188403692784\\
-0.155578022162136	-0.977411543038528\\
0.00221870262576314	0.0543990084897849\\
-0.626202961635173	-0.789034309525662\\
-0.19555812194586	-0.0447549335212659\\
-0.638261727241423	0.906066390582364\\
-0.522723531655277	0.47297099114778\\
0.89093860279797	-0.626570692626458\\
-0.379806218592581	0.870953237566705\\
-0.281428192636078	-0.89099596346109\\
0.630691571502477	-0.717932058112748\\
0.38681417481047	-0.00631054843015418\\
-1.15852520626445	0.200944140021088\\
-0.0241177568961741	0.148650176878417\\
0.365462887167067	-0.946294629672133\\
-0.056988715870508	0.366382684411122\\
0.482816362793881	-1.07179955979817\\
-0.336954106808649	-0.67607593835338\\
0.349709580238228	-0.21945055537668\\
0.769112425765599	-0.810915193201116\\
0.776125940995956	-0.351619173943707\\
0.721775638741493	0.142006094307121\\
-0.450592346219716	-0.327314578070295\\
-0.337603282178201	-0.0991101990381705\\
0.180569655256341	0.78934469648117\\
-0.436504353910743	-0.199946689589212\\
-1.01613119922048	0.529006847518921\\
-0.717353159251196	-0.677345748668364\\
0.497849574964129	-0.469462176368483\\
-0.276563227687468	1.08774432607757\\
-0.0356843841085204	-1.06355394223276\\
-0.048540938573943	1.07890177086046\\
-0.993703323617101	-0.027317762499229\\
1.01496343216943	0.0329828964562181\\
-0.999117210043507	0.288548804185077\\
0.406349858300215	-0.307840863181433\\
0.550496724693491	-0.602584696934722\\
-1.12965303066137	0.370633214571804\\
-0.61957703104798	0.693282954655554\\
-0.40616174234698	0.629367281465831\\
0.120404085276531	-1.14609223629567\\
-0.158745080458876	0.844055500568309\\
-0.689858891592535	-0.106835779714896\\
1.12436579651219	-0.142339824285221\\
-0.685675000636591	0.374470171261513\\
-0.902069028329595	-0.592574026303776\\
0.837322382287956	-0.429559777004366\\
0.576968624982754	0.650774981605874\\
0.00346356784651152	-0.221309087532338\\
0.188638881041039	-0.714019582703975\\
-0.508011716578569	0.855396290744657\\
-0.270838158138972	1.07498519613779\\
0.867603028499873	-0.276290632584587\\
0.775587937997048	0.760082939981344\\
0.413323737374699	0.492069582765337\\
0.841082148399436	-0.819706039030625\\
0.215200245583618	0.372662008455956\\
1.03355924640995	-0.584489176005439\\
1.14207974781137	-0.0392555514239157\\
-0.976308415712018	0.541791467076531\\
-0.483166511257988	0.339783073312784\\
-0.105093204921728	0.560156344675861\\
0.188909930366524	-0.575557346814142\\
-0.670397205793711	0.481397572781984\\
-0.907611640235624	-0.675281431465868\\
-0.088201304490274	-0.712812203346683\\
1.04475757173906	0.16039753283156\\
1.10774066149447	0.0149352142159627\\
0.0573057166981726	-0.910879651733436\\
1.05429805486808	0.122949334743225\\
-0.313532984201912	-0.318126513249198\\
0.802485359639796	0.134755667835597\\
0.0222818756124931	1.13348902982504\\
-0.979401304458024	0.679313118678234\\
0.273196336520798	-0.523668834853756\\
0.721765554858522	-0.523489073143217\\
-0.278855634221841	0.921400159020096\\
-0.750356346174947	-0.188272210522335\\
-0.307261368188663	-1.05056898746206\\
-1.01073401741101	0.229608235870323\\
0.379156420079824	0.676221813313718\\
0.422917672587817	-0.246261751660349\\
-0.526745825443093	0.732901278929082\\
0.209922599575218	-0.762720465462434\\
0.567760337976204	0.853715317025794\\
-0.53338481016728	-0.618127252339302\\
-0.745851961186833	-0.875986765096522\\
0.0868020818219712	-0.765422606206406\\
0.257153884150751	-0.44393762724227\\
0.5289137662676	-0.900233040967365\\
-0.770932097092996	-0.0301369767669063\\
-0.389061161236764	0.790552043664797\\
1.04113542065908	0.111229905441596\\
0.173027269290706	-0.195483940676236\\
0.565434767990955	-0.516143267890211\\
-0.717758363556229	0.592377833992863\\
0.791726324069389	-0.526631375135484\\
-0.688415966380305	-0.899177150329593\\
0.12240914929225	0.377009784914304\\
-0.342350572539344	0.00143306648658948\\
-0.0720988239728789	0.179161043168531\\
0.684309224391768	0.444650564680878\\
0.75708835332364	0.500901297744827\\
-0.71295935196991	-0.490771090047263\\
-0.0672893860877056	-0.960434614416926\\
0.0440910512274033	-0.603536097273605\\
0.72825070259317	-0.712411343187213\\
1.17374573155893	0.145931323727726\\
0.222533810803243	-0.294191566876853\\
0.128902122037027	-0.000925227646592764\\
-0.604668611407373	0.0939262702896981\\
0.247646705900901	-0.69966416690132\\
0.18550076947361	-1.11858600471167\\
-0.245755272548173	-0.583325023298942\\
0.646388758553496	0.353562914269559\\
-0.514245750544963	0.20276341614652\\
0.194132518243712	0.0173436763019608\\
-0.365934504068798	0.540314447043946\\
0.508504482181533	-0.295682285136468\\
-0.999846284536178	-0.402486093733202\\
0.389957026477588	0.607664619449863\\
0.279520114505469	0.515698066487025\\
0.871642800460132	-0.0765303792840105\\
-0.333812119342915	-0.729418252656089\\
-0.122504933320309	0.253887146182063\\
-0.0441523371291989	0.085900385351207\\
1.01084229640246	-0.322217761834229\\
-0.414528185882428	0.59638425751347\\
-0.542763654860884	-0.65606105782185\\
-0.930979626936834	-0.687793169769623\\
-0.232262220538544	0.212968122791197\\
-0.157224420606123	-0.681574953260564\\
-0.751262465851332	0.429429682275105\\
0.257321640984626	-1.15609276042909\\
1.00638548665682	-0.112066518146195\\
0.482613898520799	0.044025214817789\\
-0.145628880434107	0.730789956216359\\
-0.12407104423755	0.805547605078562\\
0.579891830717762	0.201399859652675\\
0.0832923971503158	-0.980046772009719\\
-1.1500464422926	-0.128392905039986\\
-0.486174453449418	-0.685651819903403\\
-0.0462633878208987	-1.05456717381957\\
-0.500590065760578	0.786950855860497\\
-0.462214749222301	-0.0347535162335391\\
0.24572494705723	1.16453705668631\\
-0.014220250944026	-0.882154128961715\\
0.563139916346516	0.13811384178043\\
-0.03832880705913	0.452353436474747\\
-0.609632943717996	0.230010359273034\\
1.1522455868395	-0.204190682449582\\
-0.508021965537311	-1.05539064517592\\
0.204247268802849	-0.834158327668596\\ };
\addlegendentry{Users};
	
\draw[black,thick] (150,150) circle (3.1cm);
	
\end{axis}
\end{tikzpicture}
\caption{Network geometry for the numerical example.}
\label{fig:network}
\end{figure}

\section{Simulation Results}
\label{sec:Result}
In this section, we illustrate the usefulness of the proposed optimization method with the help of following example. We considered a network geometry as shown in Fig. \ref{fig:network}. The distance between adjacent BSs is fixed to 0.8 km such that 7 BSs form an equilateral triangular lattice ($L=7$). We assume that 200 users are uniformly distributed within a circular network of radius 0.8 km and in each time slot 12 of them are served ($M=12$). The target SINR of each user is $\gamma_{m}=10$ dB, $\forall m$. Each BS has two antennas ($N_{t}=2$) and each user has a single antenna. Each BS has the same cache size $S$. The channel vector from the $l$-th BS to the $m$-th user is given by \cite{channel},
\begin{eqnarray}
\mathbf{h}_{l,m}= \mathbf{g}_{l,m} \sqrt{10^{-\frac{\rho_{l,m}}{10}}
\varphi \zeta_{l,m}}
\end{eqnarray}
where $\varphi=10$ dBi is the transmit antenna power gain of each BS, $\rho_{l,m}$ is the path loss at the distance $d_{l,m}$, $\zeta_{l,m}$ is the log-normal coefficient --- with a 0 dB mean and standard deviation of 8 dB --- that models the large-scale fading (shadowing), and  $\mathbf{g}_{l,m}$ is the complex Gaussian coefficient with variance 1 that models the small-scale fading. The channel vector changes independently at each time slot. For the path loss, the 3GPP Long Term Evolution (LTE) path loss model \cite{3GPP} with a path-loss exponent of 3.76 is used
\begin{eqnarray}
\rho_{l,m}=148.1+37.6 \logten(d_{l,m})
\end{eqnarray}
where $d_{l,m}$ is the distance between the $l$-th BS and $m$-th user in kms. All users are subject to white Gaussian noise with power spectral density -172 dBm/Hz over 10 MHz bandwidth. The total number of files available in the  cloud  is 20 ($F=20$). We investigate the costs for both coded and uncoded caching for different cache sizes. As mentioned in the introduction, in the case of uncoded caching only complete files are stored at the BSs; and, thus each BS is able to store $S$ ($BS/B=S$) files. As opposed to the uncoded caching case, in the coded caching setting, we assume that $0.5B$ parity bits of a particular file is stored in the caches of the BSs. This implies that, each BS stores the parity bits of $2S$ ($BS/0.5B=2S$) files. For the simulation, the caches of the BSs are filled with the $S$ most popular files for the uncoded caching and parity bits of the $2S$ most popular files for the coded caching, respectively. Note that each BS stores distinct parity bits of a particular file. The Zipf distribution parameter $\alpha$ is chosen as $1.2$ in the simulations. The optimization software tool CVX~\cite{cvx} is used to obtain the optimum beamforming vectors and the results are obtained by averaging over 100 time slots.

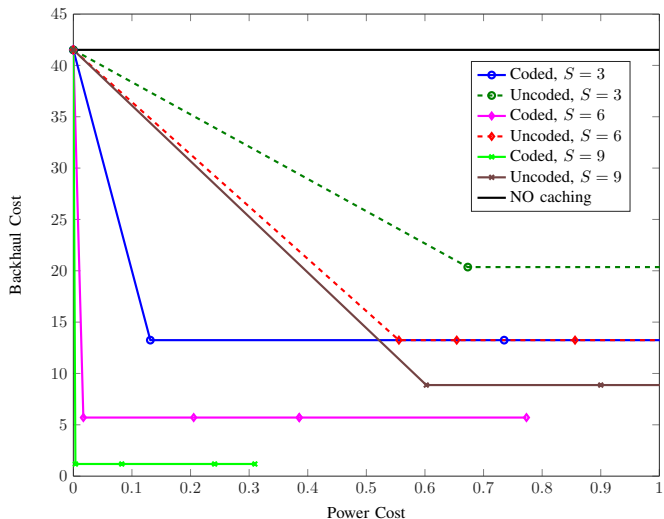
\begin{figure}[t!]
\definecolor{mycolor1}{rgb}{0.00000,0.49804,0.00000}
\definecolor{mycolor2}{rgb}{1.00000,0.00000,1.00000}
\definecolor{mycolor3}{rgb}{0.45098,0.26275,0.26275}
\begin{tikzpicture}[scale=0.628]
\begin{axis}[
width=4.88411458333333in,
height=3.85427083333333in,
scale only axis,
separate axis lines,
every outer x axis line/.append style={white!15!black},
every x tick label/.append style={font=\color{white!15!black}},
xmin=0,
xmax=1,
xlabel={Power Cost},
every outer y axis line/.append style={white!15!black},
every y tick label/.append style={font=\color{white!15!black}},
ymin=0,
ymax=45,
ylabel={Backhaul Cost},
legend style={at={(0.678787882580901,0.572386351332429)},anchor=south
west,draw=white!15!black,fill=white,legend cell align=left} ]

\addplot[color=blue,solid,mark=o,mark options={solid},line width=1.3pt]
table[row sep=crcr]{
1.9361610465644e-08	41.5131794236477\\
0.131388774553828	13.2434047094082\\
0.735157649977092	13.2434047094082\\
1.26782184888492	13.2434047094082\\
1.76879685640572	13.2434047094082\\
5.05614880013436	13.2434047094082\\ };
\addlegendentry{Coded, $S=3$};

\addplot[color=mycolor1,dashed,mark=o,mark options={solid},line width=1.3pt]
table[row sep=crcr]{
6.06943204571308e-09	41.5131794236477\\
0.673063536519982	20.3589262361215\\
2.34450415948689	20.3589262361215\\
3.10073754571152	20.3589262361215\\
3.79034361522158	20.3589262361215\\ };
\addlegendentry{Uncoded, $S=3$};

\addplot[color=mycolor2,solid,mark=diamond,mark options={solid},line width=1.3pt]
table[row sep=crcr]{
7.86401830146592e-10	41.5131794236477\\
2.9141855915661e-08	41.5131794236477\\
8.97578350780572e-10	41.5131794236477\\
6.75514483488813e-09	41.5131794236477\\
2.28746546094597e-08	41.5131794236477\\
0.0171031697086156	5.70621339759134\\
0.20537846303501	5.70621339759134\\
0.385505828407072	5.70621339759134\\
0.385505828407072	5.70621339759134\\
0.773214377597149	5.70621339759134\\ };
\addlegendentry{Coded, $S=6$};

\addplot[color=red,dashed,mark=diamond,mark options={solid},line width=1.3pt]
table[row sep=crcr]{
2.69013451247285e-08	41.5131794236477\\
0.555584454453396	13.2434047094082\\
0.654230095512546	13.2434047094082\\
0.856117897331268	13.2434047094082\\
1.06584734248995	13.2434047094082\\ };
\addlegendentry{Uncoded, $S=6$};

\addplot[color=green,solid,mark=x,mark options={solid},line width=1.3pt]
table[row sep=crcr]{
1.07198103576353e-08	41.5131794236477\\
0.00354763864575759	1.18423331241175\\
0.0825968324510746	1.18423331241175\\
0.241066278136641	1.18423331241175\\
0.309752625269433	1.18423331241175\\ };
\addlegendentry{Coded, $S=9$};

\addplot[color=mycolor3,solid,mark=x,mark options={solid},line width=1.3pt]
table[row sep=crcr]{
1.14884738503451e-09	41.5131794236476\\
3.60206138486869e-09	41.5131794236476\\
0.602292585837098	8.87060994878327\\
0.9	8.87060994878327\\
1.0208	8.8706\\ };
\addlegendentry{Uncoded, $S=9$};

\addplot[color=black,solid,line width=1.3pt]
table[row sep=crcr]{
0	41.5132\\
0.1	41.5132\\
0.2	41.5132\\
0.3	41.5132\\
0.4	41.5132\\
0.5	41.5132\\
0.6	41.5132\\
0.7	41.5132\\
0.8	41.5132\\
0.9	41.5132\\
1	41.5132\\ };
\addlegendentry{NO caching};

\end{axis}
\end{tikzpicture}	
\caption{Network cost trade-off.}
\label{fig:result}
\end{figure}

Fig. \ref{fig:result} shows the power consumption and backhaul costs trade-off for uncoded and coded caching with different cache sizes ($S=3, 6, 9$). Data points of each line are obtained by changing the weighting parameter $\lambda$. As it can be seen from Fig.~\ref{fig:result}, the backhaul cost saturates if BSs transmit with a required power.  For different cache sizes, Table \ref{table} summarizes the decrease in the backhaul costs  with the coded caching in comparison to uncoded and no caching. For sake of completeness, we also compare the uncoded caching  with no caching. We observe that coded caching always outperforms the uncoded caching cases in terms of low backhaul cost for the same cache size. This is due to the fact that since more portions of files can be stored at the BSs, files can be downloaded directly from the BSs. It is observed that the coded caching decreases the backhaul cost approximately $86\%$ compared to the uncoded caching for the cache size 9 ($S=9$). We also compare the  
benefits of the coded caching over no caching case in Table \ref{table}. By introducing a cache size of 3 ($S=3$), the backhaul cost decreases by $68.1\%$. An additional decrease of $18.1\%$ in backhaul cost is observed by doubling cache size from 3 to 6 ($S=6$). Furthermore, only $10.9\%$ additional decrease is obtained when the cache size is increased from 6 to 9 ($S=9$). A similar trend is also observed for the uncoded case. Introducing even a small size of storage at the BSs cause a substantial decrease in the backhaul cost.

\section{Conclusion}
\label{sec:Conclusion}
In this work, we consider a C-RAN model in which the BSs are connected to the cloud via finite capacity backhaul links and where local storage is also available at the BSs. We investigate this problem by minimizing the total network cost under quality of service constraints, where coded caching is allowed at the BSs.  We show that even introducing small sized caches at BSs decreases the backhaul cost significantly. In comparison to the uncoded caching strategies studied in the existing literature, we show that coded caching schemes reduce the overall network backhaul.  

\renewcommand\arraystretch{1.6}
\begin{table}[t!]
\caption{Performance Gain --- Reduced Backhaul Costs}	
\label{table}
\centering
\begin{tabular}{c|c|c|c|}
\cline{2-4}
&\begin{tabular}[c]{@{}c@{}}Coded vs.\\No caching\end{tabular} & \begin{tabular}[c]{@{}c@{}}Uncoded vs.\\ No caching\end{tabular} &\begin{tabular}[c]{@{}c@{}}Coded vs.\\Uncoded\end{tabular}\\ \hline
\multicolumn{1}{|c|}{$S=3$}&$68.1\%$&$50.9\%$&$34.9\%$\\ \hline
\multicolumn{1}{|c|}{$S=6$}&$86.2\%$&$68.1\%$&$56.9\%$\\ \hline
\multicolumn{1}{|c|}{$S=9$}&$97.1\%$&$78.6\%$&$86.6\%$\\ \hline
\end{tabular}
\end{table}

\balance
\bibliographystyle{IEEEtran}
\bibliography{yigitRefWSA}
\end{document}